\documentclass[prl,aps,twocolumn,showpacs]{revtex4}
\usepackage{epsfig}
\usepackage{bm}

\begin{document}

\title{Chaotic response of global climate to long-term solar forcing variability}

\author{\small  A. Bershadskii}
\affiliation{\small {ICAR, P.O.B. 31155, Jerusalem 91000, Israel}}

\begin{abstract}
It is shown that global climate exhibits {\it chaotic} response to solar forcing 
variability in a vast range of timescales: from annual to multi-millennium. 
Unlike linear systems, where periodic forcing leads to periodic response, 
nonlinear chaotic response to periodic forcing can result in exponentially decaying 
broad-band power spectrum with decay rate $T_e$ equal to the period of the 
forcing. It is shown that power spectrum of a reconstructed time series 
of Northern Hemisphere temperature anomaly for the past 2,000 years has 
an exponentially decaying broad-band part with $T_e \simeq 11$ yr, i.e. the observed 
decay rate $T_e$ equals the mean period of the solar activity. It is also shown that 
power spectrum of a reconstruction of atmospheric $CO_2$ time fluctuations 
for the past 650,000 years, has an exponentially decaying broad-band part with 
$T_e \simeq 41,000$ years, i.e. the observed decay rate $T_e$ equals the 
period of the obliquity periodic forcing. A possibility of a {\it chaotic} solar 
forcing of the climate has been also discussed. These results clarify role of 
solar forcing variability in long-term global climate dynamics (in particular in 
the unsolved problem of the glaciation cycles) and help in construction of adequate 
{\it dynamic} models of the global climate. 

\end{abstract}


\maketitle

\section{Introduction}

Behavior of a chaotic system can be significantly altered by applying 
of a periodic forcing. Already pioneering studies of the effect of external 
periodic forcing on the first Lorenz model of the chaotic climate 
revealed very interesting properties of chaotic response (see, for 
instance, \cite{golub},\cite{ahlers},\cite{fran}). The forcing does not always 
have the result that one might expect \cite{bg},\cite{chab},\cite{tam}. 
The climate, where the chaotic behavior was discovered for the first time, 
is still one of the most challenging areas for the chaotic response 
theory. One should discriminate between chaotic weather (time scales 
up to several weeks) and a more long-term climate variation. 
The weather chaotic behavior usually can be directly related to chaotic 
convection, while appearance of the 
chaotic properties for the long-term climate events is a non-trivial and 
challenging phenomenon. It seems that such properties can play a significant role 
even for glaciation cycles, i.e. at least at multi-millennium time scales 
\cite{hC},\cite{sal}. Cyclic forcing, due to astronomical modulations of the solar 
input, rightfully plays a central role in the long-term climate models. 
Paradoxically, it is a very non-trivial task to find imprints of this forcing in 
the long-term climate data. It will be shown in present paper that just unusual 
properties of {\it chaotic} response are the main source of this problem.  \\

\begin{figure} \vspace{-0.5cm}\centering
\epsfig{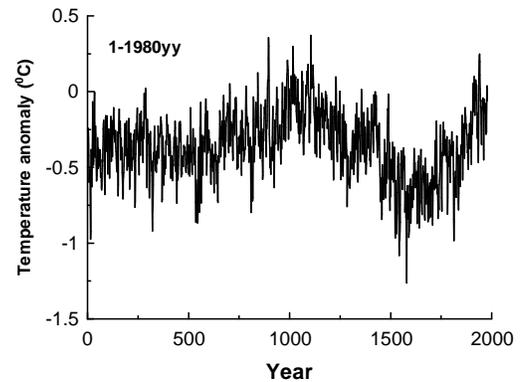} \vspace{-4.5cm}
\caption{A reconstruction of Northern Hemisphere temperature
anomaly for the past 2,000 years.}
\end{figure}

\section{Global temperature response to {\it periodic} solar forcing}

Figure 1 shows a reconstruction of Northern Hemisphere temperatures
for the past 2,000 years (the data for this figure were
taken from Ref. \cite{paleoT}). This multi-proxy reconstruction
was performed by the authors of Ref. \cite{moberg} using combination
of low-resolution proxies (lake and ocean sediments)
with comparatively high-resolution tree-ring data.
Figure 2 shows a power spectrum of the data set calculated
using the maximum entropy method, because it provides an optimal spectral
resolution even for small data sets.
The spectrum exhibits a wide peak indicating a
periodic component with a period around 22 y, and a
broad-band part with exponential decay: 
$$
E(f) \sim e^{-f/f_e}   \eqno{(1)}
$$
A semi-logarithmic
plot was used in Fig. 2 in order to
show the exponential decay more clearly (at this plot the
exponential decay corresponds to a straight line). Both
stochastic and deterministic processes can result in the
broad-band part of the spectrum, but the decay in the
spectral power is different for the two cases. The exponential
decay indicates that the broad-band spectrum
for these data arises from a deterministic rather than a
stochastic process. For a wide class of deterministic
systems a broad-band spectrum with exponential
decay is a generic feature of their chaotic solutions 
Refs. \cite{o}-\cite{hav}.

\begin{figure} \vspace{-1cm}\centering
\epsfig{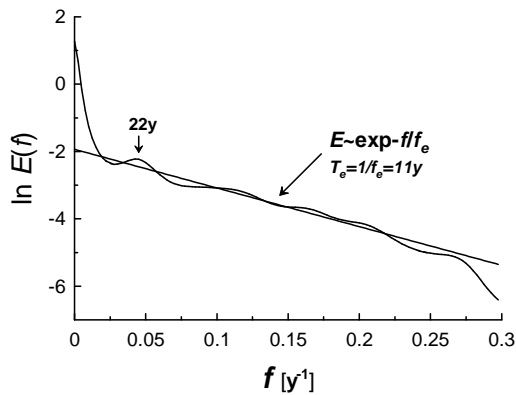} \vspace{-4.5cm}
\caption{Spectrum of the data, shown in Fig. 1, in semi-logarithmic
scales. The straight line indicates the exponential
decay Eq. (1).}
\end{figure}
\begin{figure} \vspace{-0.5cm}\centering
\epsfig{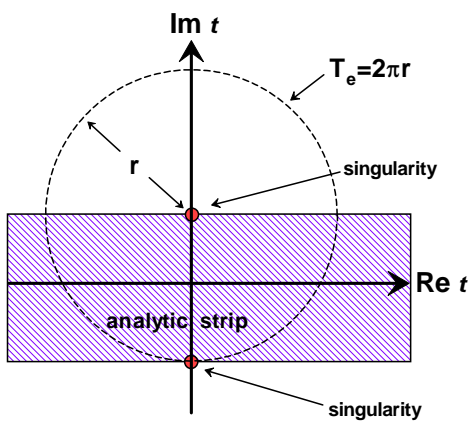} \vspace{-5cm}
\caption{A sketch of the complex time plane for the parametric modulation with period $T_e$.}
\end{figure}

Nature of the exponential decay of the power spectra
of the chaotic systems is still an unsolved mathematical
problem. A progress in solution of this problem
has been achieved by the use of the analytical continuation
of the equations in the complex domain (see, for 
instance, \cite{fm}). In this approach the exponential decay
of chaotic spectrum is related to a singularity in the
plane of complex time, which lies nearest to the real axis.
Distance between this singularity and the real axis determines
the rate of the exponential decay. For many interesting cases 
chaotic solutions are analytic in a finite strip around the real time axis. 
This takes place, for instance for attractors bounded in the real 
domain (the Lorentz attractor, for instance). 
In this case the radius of convergence of the Taylor series 
is also bounded (uniformly) at any real time. 
If parameters of the dynamical system fluctuate periodically 
around their mean values with period $T_e$, then the restriction of the Taylor series 
convergence (at certain conditions) is determined by the period of the 
parametric modulation, and the width of the analytic strip around 
real time axis equals $T_e/2\pi$ (cf. Fig. 4). Let us consider, for 
simplicity, solution $u(t)$ with simple poles only, and to define the Fourier 
transform as follows
$$
u(\omega) =(2\pi)^{-1/2} \int_{-T_e/2}^{T_e/2} dt~e^{-i \omega t} u(t)  \eqno{(2)}
$$  
Then using the theorem of residues
$$
u(\omega) =i (2\pi)^{1/2} \sum_j R_j \exp (i \omega x_j -|\omega y_j|)  \eqno{(3)}
$$
where $R_j$ are the poles residue and $x_j + iy_j$ are their location in the relevant half
plane, one obtains asymptotic behavior of the spectrum $E(\omega)= |u(\omega)|^2$ at large $\omega$
$$
E(\omega) \sim \exp (-2|\omega y_{min}|)  \eqno{(4)}
$$
where $y_{min}$ is the imaginary part of the location of
the pole which lies nearest to the real axis. Therefore, exponential
decay rate of the broad-band part of the system
spectrum, Eq. (1), equals the period of the parametric forcing.

The chaotic spectrum provides two different characteristic
time-scales for the system: a period corresponding to
fundamental frequency of the system, $T_{fun}$, and a period
corresponding to the exponential decay rate, $T_e = 1/f_e$
(cf. Eq. (1)). The fundamental period $T_{fun}$ can be estimated
using position of the low-frequency peak, while the
exponential decay rate period $T_e = 1/f_e$ can be estimated
using the slope of the straight line of the broad-band part
of the spectrum in the semi-logarithmic representation
(Fig. 2). From Fig. 2 we obtain $T_{fun} \simeq 22\pm 2$y
and $T_e \simeq 11 \pm 1y$ (the estimated errors are statistical
ones). Thus, the solar activity period of 11 years
is really a dominating factor in the chaotic temperature
fluctuations at the annual time scales, although it is
hidden for linear interpretation of the power spectrum. 
In the nonlinear interpretation the additional
period $T_{fun} \simeq 22y$ might correspond to the fundamental
frequency of the underlying nonlinear dynamical system.
It is surprising that this period is close to the 22y period
of the Sun's magnetic poles polarity switching. It should be noted 
that the authors of Ref. \cite{mu} found a persistent 
22y cyclicity in sunspot activity, presumably related to 
interaction between the 22y period of magnetic poles polarity switching 
and a relic solar (dipole) magnetic field. Therefore, one cannot rule out 
a possibility that the broad peak, in a vicinity of frequency corresponding to the 22y period, 
is a quasi{\it-linear} response of the global temperature to the weak periodic 
modulation by the 22y cyclicity in sunspot activity. I.e. strong enough 
periodic forcing results in the non-linear (chaotic) response whereas a weak periodic component of the 
forcing can result in a quasi-linear (periodic) response.

\begin{figure} \vspace{-1cm}\centering
\epsfig{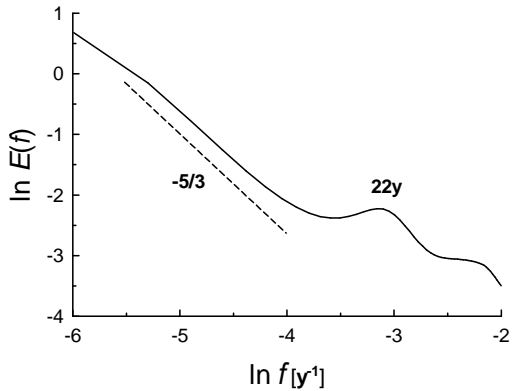} \vspace{-4.5cm}
\caption{Spectrum of the data, shown in Fig. 1, in ln-ln
scales. The dashed straight line indicates the Kolmogorov-like
spectrum: $E(f)\sim f^{-5/3}$. The high-frequency part has
been cut in order to show the low-frequency part.}
\end{figure}
\begin{figure} \vspace{-0.5cm}\centering
\epsfig{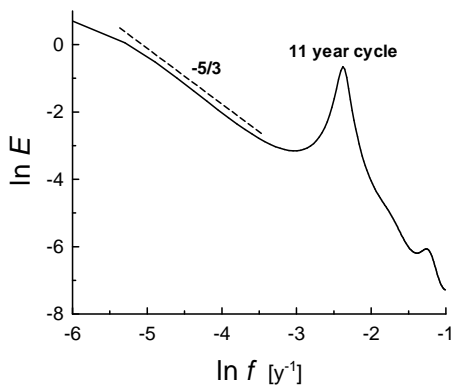} \vspace{-4.5cm}
\caption{Spectrum of the galactic cosmic ray count rate fluctuations 
in the ln-ln scales. The straight dashed line 
is drawn to indicate the Kolmogorov scaling law $E(f) \sim f^{(-5/3)}$.}
\end{figure}

Figure 4 shows the same power spectrum as in 
Fig. 2 but in ln-ln scales. A dashed 
straight line in this figure indicates a scaling with '-5/3' exponent: $E(f) \sim f^{-5/3}$. 
Although, the scaling interval is short, the value of the exponent is rather intriguing. 
This exponent is well known in the theory of fluid (plasma) turbulence and corresponds to so-called 
Kolmogorov's cascade process. This process is very universal for turbulent 
fluids and plasmas \cite{gibson},\cite{clv}. For turbulent processes on Earth and in Heliosphere 
the Kolmogorov-like spectra with such large time scales cannot exist. Therefore, one should think about 
a Galactic origin of Kolmogorov turbulence (or turbulence-like processes \cite{gibson1})
with such large time scales. This is not surprising if we recall possible role of the galactic cosmic rays for Earth climate (see, for instance, \cite{uk}-\cite{b2}). In order to support this point 
we show in figure 5 spectrum of galactic cosmic ray intensity at the Earth's orbit (reconstruction for period 1611-2007yy \cite{usos-recon}, cf. also Ref. \cite{b1}). One can compare Fig. 5 with Fig. 4 corresponding 
to the global temperature anomaly fluctuations. It should be noted, that the above discussed response of the 
global temperature to the solar activity cycles can also have a cosmic rays variability as a transmitting agent. Indeed, the interplanetary magnetic field strongly interacts with the cosmic rays. Therefore, the change
in the interplanetary magnetic field intensity, due to the solar activity changes, can affect the Earth 
climate through the change of the cosmic rays intensity and composition 
(see Refs. \cite{uk}-\cite{b2}).

\section{Long-term temperature response to {\it chaotic} solar forcing}

It is interesting, that the solar activity itself is 
chaotic at the {\it multi}-decadal time-scales \cite{b}. 
Figure 6 shows a spectrum of a long-range reconstruction of the
sunspot number fluctuations for the last 11,000 years
(the data, used for computation of the spectrum, is available
at \cite{ssn-recon}). The semi-logarithmic scales and the straight line 
are used in this figure to indicate the chaotic nature of the sunspots 
number fluctuations at multi-decadal time scales. The slope of the straight 
line provides us with the characteristic {\it chaotic} time scale 
$T_e \simeq 176$y  for the chaotic solar activity fluctuations \cite{b}. 
It should be noted that the 176y
period is the third doubling of the fundamental solar period 22y, 
which we have discussed above (see also \cite{b} and references therein). 
As one can conclude from Fig. 4 the galactic turbulence 
has more profound effect on the global temperature than the solar forcing 
variability at these time-scales, at least for the two last millenniums. 
However, turbulence is a highly {\it intermittent} phenomenon (see, for instance, 
review \cite{sa}). Therefore, during the last two millenniums (presented in the Fig. 4) 
the solar system could pass through a very intensive galactic turbulence patch, 
while before this (and after this) the nearest galactic environment could be rather 
quiet. This can allow us to investigate the question: Whether {\it chaotic} parametric 
modulation of a nonlinear system (in our case the Earth climate) also results in the 
chaotic response (in previous Chapter we have studied {\it periodic} parametric 
modulation of this system). Figure 7 shows a reconstruction of Antarctic temperature
for the past 10,000 years using a high-resolution deuterium data set at EPICA Dome C Ice Core 
(the data for this figure were taken from Ref. \cite{EPICA}). 
Figure 8 shows spectrum corresponding to these data. The slope of the straight line provides 
us with $T_e \simeq 176$y (cf Fig. 6). 
Thus one can conclude that in this case we observe the chaotic response to the parametric modulation 
of the climate by the chaotic fluctuations of the solar activity.

\begin{figure} \vspace{-0.5cm}\centering
\epsfig{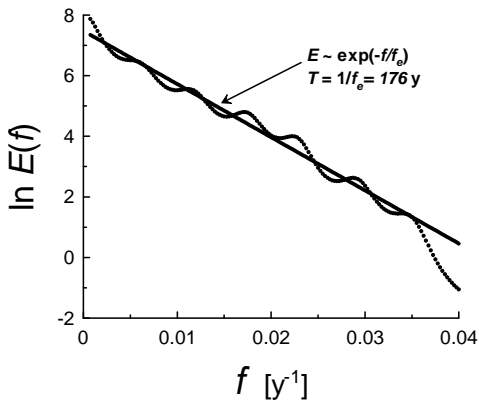} \vspace{-5cm}
\caption{Spectrum of the sunspots number fluctuations in the
semi-logarithmic scales for the reconstructed data for the last 11000 years
(the data have been taken from Ref. \cite{ssn-recon}). The straight line is drawn to
indicate the exponential law Eq. (1).}
\end{figure}

\begin{figure} \vspace{-0.5cm}\centering
\epsfig{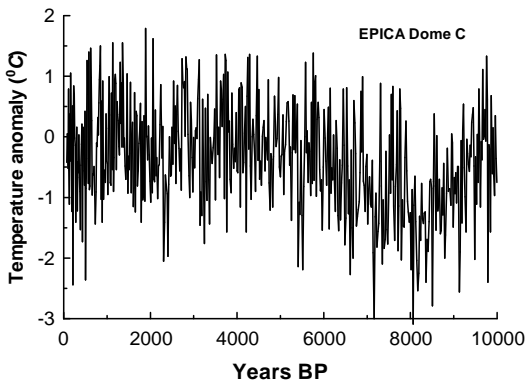} \vspace{-4.5cm}
\caption{A reconstruction of Antarctic temperature
for the past 10,000 years (EPICA Dome C Ice Core  \cite{EPICA}).}
\end{figure}
\begin{figure} \vspace{-1cm}\centering
\epsfig{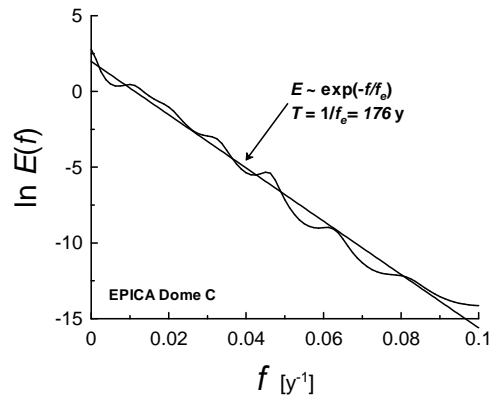} \vspace{-4cm}
\caption{Spectrum of the data, shown in Fig. 7, in semi-logarithmic
scales. The straight line indicates the exponential
decay Eq. (1).}
\end{figure}

\section{The problem of glaciation cycles}

The angle between Earth's rotational axis and the normal to the plane of its orbit 
(known as {\it obliquity}) varies periodically between 22.1 degrees and 24.5 degrees 
on about 41,000-year cycle. Such  multi-millennium timescale changes in orientation change 
the amount of solar radiation reaching the Earth in different latitudes. 
In high latitudes the annual mean insolation (incident solar radiation) decreases with 
obliquity, while it increases in lower latitudes. Obliquity forcing effect is maximum 
at the poles and comparatively small in the tropics. 
Milankovi\'{c} theory suggests that lower obliquity, leading to reduction in summer insolation 
and in the mean insolation in high latitudes, favors gradual accumulation of ice and snow
leading to formation of an ice sheet \cite{sal}. The obliquity forcing on Earth 
climate is considered as the primary driving force for the cycles of glaciation (see for a recent 
review \cite{rh}). Observations show that glacial changes from -1.5 to -2.5 Myr (early Pleistocene) 
were dominated by 
41 kyr cycle \cite{hC},\cite{rm},\cite{h1}, whereas the period from 0.8 Myr to present (late Pleistocene)
is characterized by approximately 100 kyr glacial cycles \cite{hays},\cite{im}. 
While the 41 kyr cycle of early Pleistocene glaciation is readily related to the 41 kyr period 
of Earth's obliquity variations 
the 100 kyr period of the glacial cycles in late Pleistocene still presents a serious problem. 
Influence of the obliquity variations on global climate started amplifying around 2.5 Myr,    
and became nonlinear at the late Pleistocene. Long term decrease in atmospheric $CO_2$, which 
could result in a change in the internal response of the global carbon cycle to the obliquity forcing, 
has been mentioned as one of the principal reasons for this phenomenon (see, for instance, 
\cite{berg1}-\cite{clar}). Therefore, investigation of the historic variability in 
atmospheric $CO_2$ can be crucial for understanding the global climate changes at millennial 
timescales. 
\begin{figure} \vspace{-1cm}\centering
\epsfig{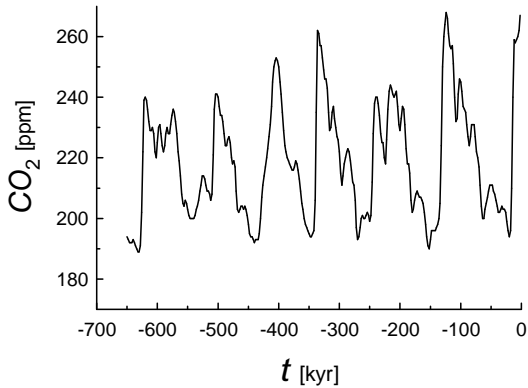} \vspace{-4.5cm}
\caption{A reconstruction of atmospheric $CO_2$ based on deep-sea proxies, 
for the past 650kyr. }
\end{figure}
\begin{figure} \vspace{-0.5cm}\centering
\epsfig{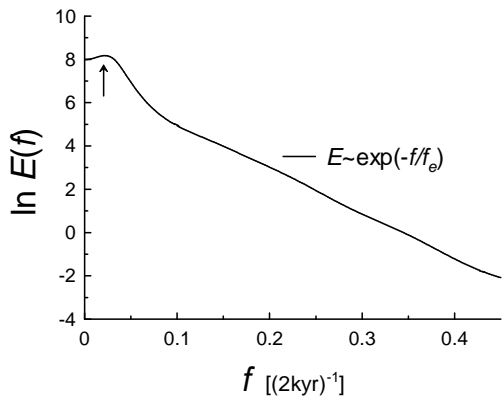} \vspace{-4.5cm}
\caption{Spectrum of atmospheric $CO_2$ fluctuations for the data shown in Fig. 9}
\end{figure}
 Figure 9 shows a reconstruction of atmospheric $CO_2$ based on deep-sea proxies, 
for the past 650kyr (the data taken from \cite{berg-data}). Resolution of the data set is 
2kyr. Fluctuations with time-scales less than 2kyr could be rather large 
(statistically up to 308ppm \cite{berg-data}), but they are smoothed by the resolution. 
Figure 10 shows a power spectrum of the data set calculated using the maximum entropy method. 
The spectrum exhibits a peak indicating a periodic component 
(the arrow in the Fig. 10 indicates a 100kyr period) and a broad-band part with exponential decay. 
A semi-logarithmic plot was used in Fig. 10 in order to show the exponential decay more clearly 
(at this plot the exponential decay corresponds to a straight line).  From Fig. 10 we obtain 
$T_{fun} \simeq 95 \pm 8$ kyr (the peak is quite broad 
due to small data set) and $T_e \simeq 41 \pm 1$ kyr 
(the estimated errors are statistical ones). Thus, the obliquity period of 41 kyr is 
still a dominating factor in the chaotic $CO_2$ 
fluctuations, although it is hidden for linear interpretation of the power spectrum. 
In the nonlinear interpretation the additional period $T_{fun}\simeq 100$ kyr might correspond 
to the fundamental frequency of the underlying nonlinear dynamical system and it determines the 
apparent 100 kyr 'periodicity' of the glaciation cycles for the last 650 kyr 
(cf Refs. \cite{sal},\cite{ru},\cite{hug1} and references therein). And again, 
as in the above considered case of the global temperature fluctuations, one cannot rule out 
a possibility that the broad peak, in a vicinity of frequency corresponding to the 100 kyr period, 
is a quasi{\it-linear} response of the atmospheric $CO_2$ to the weak periodic 
modulation by the 100 kyr cyclicity in the orbital eccentricity variations \cite{sha}. I.e. again, 
strong enough periodic forcing results in the non-linear (chaotic) response whereas a weak periodic component of the forcing can result in a quasi-linear (periodic) response. \\
 
The author is grateful to W.H. Berger, to V. Masson-Delmotteto, to A. Moberg, to J.J. Niemela and 
I.G. Usoskin for sharing their data and discussions.

\end{document}